\documentclass[compsoc,conference,a4paper,10pt,times]{IEEEtran}
\IEEEoverridecommandlockouts
\usepackage{cite}
\usepackage{amsmath,amssymb,amsfonts}
\usepackage{algorithmic}
\usepackage{graphicx}
\usepackage{textcomp}
\usepackage{bmpsize}
\usepackage{amsmath}
\usepackage{multirow}
\usepackage{xcolor}
\usepackage{lipsum}
\usepackage[colorlinks=true,urlcolor=black]{hyperref}
\def\BibTeX{{\rm B\kern-.05em{\sc i\kern-.025em b}\kern-.08em
    T\kern-.1667em\lower.7ex\hbox{E}\kern-.125emX}}
\begin{document}
\author{\IEEEauthorblockN{Shova Kuikel}
	\IEEEauthorblockA{University of Texas at El Paso\\
		skuikel@miners.utep.edu}
	\and
	\IEEEauthorblockN{Aritran Piplai}
	\IEEEauthorblockA{University of Texas at El Paso\\
		apiplai@utep.edu}
	\and
	\IEEEauthorblockN{Palvi Aggarwal}
	\IEEEauthorblockA{University of Texas at El Paso\\
		paggarwal@utep.edu}}
\title{{Evaluating Large Language Models for Phishing Detection, Self-Consistency, Faithfulness, and Explainability}} 

\maketitle

\begin{abstract}

Phishing attacks remain one of the most prevalent and persistent cybersecurity threat with attackers continuously evolving  and intensifying tactics to evade the general detection system. Despite significant advances in artificial intelligence and machine learning, faithfully reproducing the interpretable  reasoning with classification and explainability that underpin phishing judgments remains challenging. Due to recent advancement in Natural Language Processing, Large Language Models (LLMs) show a promising direction and potential for improving domain specific  phishing classification tasks. However, enhancing the reliability and robustness of classification models requires not only accurate predictions from LLMs but also consistent and trustworthy explanations aligning with those predictions. Therefore, a key question remains: Can LLMs not only classify phishing emails accurately but also generate explanations that are reliably aligned with their predictions and internally self-consistent? To answer these questions, we have fine-tuned transformer-based models, including BERT, Llama models, and Wizard, to improve domain relevance and make them more tailored to phishing specific distinctions, using Binary Sequence Classification, Contrastive Learning (CL) and Direct Preference Optimization (DPO). To that end, we examined their performance in phishing classification and  explainability by applying the ConsistenCy measure based on SHAPley values (CC-SHAP), which measures prediction–explanation token alignment to test the model’s internal faithfulness and consistency and uncover the rationale behind its predictions and reasoning. Overall, our findings show that Llama models exhibit stronger prediction–explanation token alignment with higher CC-SHAP scores despite lacking reliable decision‐making accuracy, whereas Wizard achieves better prediction accuracy but lower CC-SHAP scores. The code in available in the GitHub repository \footnote{https://github.com/PsyberSecLab/Fine-Tuning-and-Explainability-for-Phishing-Detection}.


\end{abstract}


\section{Introduction}

Phishing attacks are malicious activities where attackers try to impersonate themselves as a trustworthy source to get sensitive information from certain targeted users \cite{singh2019training}. Despite the massive technological shift over cybersecurity over several years, many organizations, individuals, and even security experts are becoming victims of phishing emails. It has been reported by the Anti-Phishing Working Group (APWG) that the number of phishing emails exponentially grew from\textbf{ 44,008 in first quarter of 2020 to 128,926 by the end of the year} \cite{salloum2022systematic}. The techniques such as rule-based detection, email filters and blacklisting of malicious domains served as a traditional anti-phishing measures. However, the dynamic tactics of attackers like obfuscation, impersonation, and personalized social engineering methodologies bypassed the conventional strategies for defense \cite{dhamija2006phishing},highlighting the need for more advanced and sophisticated detection methods.

To address this challenge, initially, scholars were focused towards Machine Learning methods with complex and variety of dataset to comprehend the complex patterns within emails which is often considered as indicative measures in phishing attempt and detection \cite{do2022deep}. These models utilize the mathematical principles of weighted feature analysis to make predictions. However, despite their effectiveness in many cases in various domain, they often struggle to generalize across evolving attack strategies, lack of interpretability and failing to capture the nuanced reasoning behind the deceptive  tactics commonly seen in phishing domain. 

With the advancement of Large Language Models LLMs, researchers have redirected their studies towards innovative approaches in phishing detection \cite{koide2024chatspamdetector} and analysis. The usage of LLMs includes developing detection systems \cite{heiding2024devising}\cite{lee2025enhancing} \cite{jamal2024improved}, creating LLM-generated emails to assess the resilience of current phishing detection tools \cite{hazell2023spear} \cite{afane2024next} and conducting experimental studies to determine human susceptibility to LLM-crafted phishing emails \cite{sharma2023well}. Additionally, LLMs are used in explainable AI (XAI) to translate technical outputs into natural language explanations, as in EXPLICATE \cite{lim2025explicate}, enhancing user understanding and trust in phishing detection systems. Likewise, a fine-tuned models were used for phishing detection, where explainability was achieved through LIME and Transformer Interpret techniques \cite{uddin2024explainable}. Even though they demonstrated the effectiveness of integrating explainability into phishing detection, they primarily rely on post-hoc methods where predictions are first generated by an ML model and then interpreted separately. 

 However, this area of research is still fundamentally under-explored and limited. In this research study, we aim to systematically investigate different potential of LLMs in phishing detection to explore its capabilities in  classification, explanability, and self-consistency to bridge the gap and develop a nuanced understanding of LLMs' potential in cybersecurity. To achieve this goal from this research study, we want to address the following interconnected research questions: 1) How accurate LLMs are while performing phishing classification tasks?, The mere accuracy of these models may not be enough to make LLM models reliable and faithful, thus, we explore the next question i.e. 2) Can LLMs provide internally consistent and contextually grounded explanations that align with their predictions ?

\section{Dataset description}
We collected the experimental dataset for this research from various sources, as obtaining high-quality datasets for phishing detection is a challenging task itself. Out of three datasets, the \textbf{ Enron Email Corpus}, a real-world dataset comprising of 500,000 emails gathered from 158 employees of the Enron corporation, was the primary source for our ham email dataset \cite{bountakas2021comparison}. Similarly, we used the \textbf{  Nazario } dataset for phishing emails \cite{nazario2021}, which comprised of phishing emails collected between 2004 to 2020 \cite{chanis2025enhancing}. We used the phishing emails from 2015-2024 as we were only focused on certain attributes of an email such as newer phishing tactics and strategies that have evolved over recent years.  As the Nazario dataset collects a large number of phishing emails over a considerable period of time, it is highly respected and often used in phishing-related research. These two datasets were used to fine-tune our various LLMs for phishing detection, and a subset of the same data was also utilized to evaluate the CC-SHAP method.

 \begin{table}[h!]
\centering
\begin{tabular}{|l|l|l|l|}
\hline
\textbf{Source}         & \textbf{Email Type} & \textbf{Count}         \\ \hline
Nazario  \cite{nazario2021}               & Phishing           & 2500                       \\ \hline
Enron  \cite{bountakas2021comparison}                 & Ham                & 2500                      \\ \hline
\end{tabular}

\caption{Dataset Description}
\label{tab:datasets}
\end{table}

In order to ensure cleanliness, consistency, and suitability, the dataset we obtained underwent intensive data pre-processing steps before actual training. The following are the pre-processing steps that we applied on the dataset:
 
\subsection{ Data Cleaning} 
We removed the HTML tags, special symbols, emails that were in different languages, decoded ASCII codes, and unnecessary characters. We used the BeautifulSoup library to parse the HTML email content to remove noise and concentrate on extracting the necessary text from the email.

\subsection{Duplicate Removal Balanced Dataset}
To avoid over-representation and maintain the dataset's diversity, redundant entries were identified and removed.To balance the dataset and to ensure the model does not bias towards any class of emails, we took the same number of email counts (2500) from both Nazario :phishing and Enron :ham for fine-tuning purposes.

\subsection{Feature Extraction}
Among multiple attributes and characteristics of an email present in dataset, we were interested in certain attributes of an email such as: \textbf{Body}, \textbf{Sender}, \textbf{Subject} and \textbf{Label}. Here, our objective was to find the nuanced semantic and contextual patterns in these extracted features to distinguish between two types of emails that eventually served as a distinct purpose in fine-tuning and evaluation tasks.

\section{Methodology}
Transformer architecture-based LLMs are outstanding at understanding complex linguistic patterns and relationships, as well as text generation, as they are pre-trained on massive corpora. These abilities help them achieve the state-of-the-art performance in various Natural Language Processing tasks such as text classification, text summarization, language translation, sentiment analysis and so on. Even though these models are proficient at a wide range of Natural Language Processing tasks, their performance as well as effectiveness in domain-specific areas like cybersecurity remain a challenge with resource-intensive training and domain-specific datasets. To resolve the issues, one efficient approach is to leverage knowledge by fine tuning LLMs with high-quality domain-related data while minimizing the need of extensive pre-training on a huge number of parameters \cite{zhang2023hackmentor}.

Rather than starting from scratch, fine-tuning uses the existing knowledge base of that particular pre-trained LLM that it learned while training. During the fine-tuning process, it updates only the required parameters to adjust and fit to certain specific requirements of the tasks \cite{lin2024data}. With this underlying principle, we fine-tuned different LLM models with our cyber domain-specific dataset for binary email classification. The main purpose was to get more contextualized embeddings vectors that can effectively differentiate between two classes of emails: phishing and ham. Considering the resources required, we employed Parameter Efficient Fine-Tuning (PEFT) techniques \cite{han2024parameter} such as Low-Rank Adaptation (LoRA) which simply updates the small set of parameters without requiring to update all the parameters in pre-trained model.


   

We implemented LoRA \cite{hu2021lora} based fine-tuning of LLMs. The fundamental idea behind LoRA is to freeze the weights of the original model while adding a small subset of trainable sub-modules to train with additional network layers in transformer architecture. \cite{zhang2023hackmentor} found that fine-tuning only the query and value matrices, rather than all four attention matrices, achieves comparable performance. By keeping only query and value as target modules for LoRA, we fine-tuned our three different large language models: \textbf{Llama-2-7B}, \textbf{Llama-3-8B}, \textbf{Wizard-7B}. Additionally, we also included \textbf{BERT} model for fine tuning with general architecture as it is smaller model compared to other models.

\section{Experiments}

\subsection{Models}

In this research study, we used various language models to test different aspects of phishing email detection, including \textbf{BERT: bert-base-uncased}, \textbf{Llama 7B: meta-llama/Llama-2-7b-hf}, \textbf{Llama 8B: llama/Llama-3-8B} and \textbf{Wizard 7B: dreamgen/WizardLM-2-7B}.These models were trained on different approaches of fine-tuning to detect phishing and legitimate emails based on the features.

Bidirectional Encoder Representations from Transformers (BERT) \cite{devlin2019bert} was first introduced by Google AI in 2018,which is a ground breaking NLP model known for its bidirectional understanding of text using Masked Language Modeling (MLM) and Next Sentence Prediction (NSP).The reason for choosing BERT for this study, is for its established capabilities in capturing the nuanced relationship between words and sentences in an emails \cite{al2024comparative}. Llama 7B \cite{llama} and Llama 8B \cite{llama3_8b} are both open-source language models developed by Meta AI and released in 2023 and 2024 respectively. These transformer based models from Meta AI, were leveraged for their ability to process long sequences and their pre-training on large corpora, which is required for effective understanding of phishing strategies. Wizard 7B \cite{wizardlm2-7b}, as fine tuned derivative of foundational LLM Mistral, was chosen for its strong instruction-following abilities and advanced attention mechanism with evol-instruct tuning allowing it to capture phishing characteristics precisely.

\subsection{Fine-Tuning Approach}
We have employed three major approaches for fine tuning: 
\textbf{Binary Sequence Classification},
\textbf{Direct Preference Optimization}, and \textbf{Contrastive Learning} using different pre-trained large language models as explained in above section.

\subsubsection{Binary Sequence Classification}
As one of our fine-tuning approach, we employed general binary sequence classification for phishing detection treats each email as a sequence of tokens, often combining sender, subject, and body and uses a pretrained LLMs models to produce contextualized embeddings for accurate decision. During training, the model minimizes cross‐entropy loss on labeled examples. With the fine‐tuning on a curated dataset of phishing and ham emails, the model learns to recognize textual patterns, explicit features and semantic cues that distinguish phishing and legitimate emails.

\subsubsection{Direct Preference Optimization}

During the fine- tuning process, we used  the Nazario and Enron dataset to train our LLM models to differentiate phishing emails from the legitimate. We employed Direct Preference Optimization (DPO), a stable and computationally efficient fine-tuning method that eliminates the need for reward model fitting or extensive hyper-parameter tuning\cite{rafailov2023direct}. Rather than relying on explicit labels, our models aimed to learn from the structured email comparisons, analyzing key features such as sender, subject, and body to detect phishing patterns. DPO learns by minimizing the loss, while encouraging models to prefer email structures that align with phishing or legitimate characteristics to replicate its decision-making processes with human-like preferences. 


\subsubsection{Contrastive Learning }
We also employed Contrastive Learning approach to fine-tune the pre-trained language model for the task of phishing email detection.In Contrastive Learning based fine tuning \cite{le2020contrastive}, models learn by comparing emails in structured triplets, where one email serves as a reference, another is similar in intent, and the third is the one with dissimilar intent. The email input includes the key features of an email like sender, subject and the body. In the process of training, the model distills its understanding by pulling similar emails closer in its learned representation. While the models push different ones apart, creating a subtle distinguished detection scenario for phishing emails.



\section{Explainability and Self-Consistency of LLMs}

The explainability is crucial in phishing classification to ensure that model predictions and explanations are transparent and replicate with human-like reasoning. Quantifying natural language explanations is challenging and an emerging research area. We employ the concept of 
 \textbf{Consistency Measured based on SHAPley Values (CC-SHAP)} derived from the prior research \cite{parcalabescu2023measuring}, and adapt it for phishing classification.  In our implementation, to quantify how well the model's prediction aligns with its core reasoning and decision-making processes, we extended the original CC-SHAP methodology by computing the SHAP values from the input email text for both the classification and explanation.

The classification SHAP values measure how much each token in the email contributes to that particular decision- PHISHING or LEGITIMATE by evaluating probability shift when individual tokens are masked. 
To compute SHAP values, we used a perturbation-based masking strategy that selectively replaces tokens with the padding token and evaluates the probability shift in classification outcomes.
 Specifically, for each token j equation (\ref{equation:shap}) in the input, its contribution is approximated using Monte Carlo sampling as:

\begin{equation}
\label{equation:shap}
\phi_j \;=\; \frac{1}{N} \sum_{s \in S_j} \bigl(P\bigl(s \cup \{j\}\bigr) - P(s)\bigr)
\end{equation}
where:
\begin{itemize}
    \item $N$ is the total number of Monte Carlo samples,
    \item $S_j$ is the set of sampled token coalitions that do \emph{not} include token $j$,
    \item $P\bigl(s \cup \{j\}\bigr)$ denotes the model’s output probability when token $j$ is unmasked together with the tokens in coalition $s$,
    \item $P(s)$ denotes the model’s output probability when only the tokens in coalition $s$ are visible (all other tokens, including $j$, are masked).
\end{itemize}

During implementation, we randomly select coalitions of tokens, create masked inputs preserving only those coalition tokens, compute each token’s marginal contribution as the difference between the model’s output probability when token j is included in the coalition and the model’s output probability when only the coalition tokens are present. Then, accumulate these contributions across all samples.
These SHAP values are normalized by computing a contribution ratio for each token as:
\begin{equation}
\label{equation:Normalize}
    c_j = \frac{\phi_j}{\sum_{i} |\phi_i|}
\end{equation}
which scales the contributions to values within the range of $-1$ to $1$. The shap values are obtained for both the prediction and the explanation. Subsequently, the cc-shap score is obtained by computing the cosine distance between normalized SHAP vectors for both prediction and explanation, ensuring  both vectors are of equal length  by re-normalizing them with L1 norm.

\begin{equation}
\text{CC-SHAP} = 1 - \text{cosine-dist}(\phi^{(p)}_{\text{norm}}, \phi^{(e)}_{\text{norm}})
\end{equation}

\section{Results}
\subsection {Accuracy of LLM Models in Predicting Ground Truth in Phishing Datasets}

We evaluated  different LLM models with our combined dataset (Nazario and Enron) and compared the  models performance using three types of fine-tuning approaches such as: binary classification, contrastive learning and direct preference optimization. It is observed from Table \ref{tab:binary_classification} that binary classification based fine -tuning yielded the good performance across all the models, with BERT achieving $ 98.89$ training accuracy and $98.55$ validation accuracy. The results revealed that Llama 7B and Llama 8B also showed strong performance in binary classification with validation accuracies of $90.90$ and $93.30$ respectively. However, Wizard 7B consistently underperformed, with $83.68$ validation accuracy with higher loss of 0.98. This results suggest that Wizard 7B has the weaker generalization compared to other models.

\begin{table*}
\centering
\begin{tabular}{|l|c|c|c|c|}
\hline
\textbf{Model}  & \textbf{Training Accuracy (\%)} & \textbf{Validation Accuracy (\%)} & \textbf{Training Loss} & \textbf{Validation Loss} \\ \hline
BERT      & 98.89  & 98.55  & 0.03   & 0.04  \\ \hline
LLaMA 7B  & 92.15  & 90.90  & 0.18   & 0.19    \\ \hline
LLaMA 8B  & 94.37  & 93.30  & 0.15   & 0.19    \\ \hline
Wizard 7B & 84.68  & 83.68  & 0.87   & 0.98    \\ \hline
\end{tabular}
\caption{Binary Classification Results for Different Models Against Ground Truth of Emails}
\label{tab:binary_classification}
\end{table*}

Additionally, with an intention of learning more nuanced semantics involved in the phishing detection which can be reflected in the embeddings, we employed the contrastive based learning. This approach allowed our models to develop more refined contextual representations by comparing positive and negative examples rather than simply predicting the class labels. With this, resulting in embeddings that better reflect the semantic nuances that differentiate the legitimate emails from deceptive signal as phishing emails.The results from Table \ref{tab:contrastive_dpo} showed us BERT again has the lowest training and validation losses: $0.003$ and $0.007$ respectively. Similarly, the Llama 7B and Llama 8B performed comparably low under contrastive learning with training losses of 0.101 and 0.090 and validation losses of 0.137 and 0.088, respectively. Wizard 7B, in contrast to the LLaMA models, exhibited a training loss of 0.451 and a validation loss of 0.463, suggesting that its embeddings may be less reflective and effective in clustering phishing and legitimate emails.

\begin{table*}
\centering
\begin{tabular}{|l|l|c|c|}
\hline
\textbf{Fine-Tuning Approach} & \textbf{Model}  & \textbf{Training Loss} & \textbf{Validation Loss} \\ \hline
\multirow{4}{*}{Contrastive Learning}  
 & BERT      & 0.003  & 0.007   \\ \cline{2-4} 
 & LLaMA 7B  & 0.101  & 0.137   \\ \cline{2-4} 
 & LLaMA 8B  & 0.090  & 0.088   \\ \cline{2-4} 
 & Wizard 7B & 0.451  & 0.453   \\ \hline
\multirow{4}{*}{Direct Preference Optimization}  
 & BERT      & 0.063  & 0.068   \\ \cline{2-4} 
 & LLaMA 7B  & 1.493  & 1.683   \\ \cline{2-4} 
 & LLaMA 8B  & 1.405  & 1.563   \\ \cline{2-4} 
 & Wizard 7B & 10.98  & 11.58   \\ \hline
\end{tabular}
\caption{Contrastive Learning and Direct Preference Optimization Fine-Tuned Results}
\label{tab:contrastive_dpo}
\end{table*}
 The Direct Preference Optimization (DPO) learns a preference ordering between responses by encouraging the model to assign higher probabilities to preferred responses and lower probabilities to rejected responses \cite{rafailov2023direct}. Likewise, we employed DPO to refine the model's ability to capture the subtle preference differences, while ensuring a  enhanced understanding of phishing and ham emails. However,direct preference optimization (DPO) resulted in significantly worse performance across all models, with losses substantially higher than both the binary classification and contrastive learning. We can see from the Table \ref{tab:contrastive_dpo} that Llama 7B and Llama 8B both has training losses more than $1.4$, suggesting that they struggled to optimize preference-based ranking for phishing classification. Moreover, the Wizard 7B performed the worst beside the types specific models we used with training and validation losses of $10.98$ and $11.58$ respectively, suggesting this model is not well-suited for this training methodology.

Overall, binary classification results suggests that this is the most effective approach for phishing detection, with BERT outperforming all models in accuracy and loss reduction.

\subsection{LLM Model's Explanability, Consistency, and Faithfulness}

Building upon our previous findings, where we explored the classification performance and learning capabilities of fine-tuned model, we extended our research to test the explanability, consistency, and faithfulness of these models. To assess model performance, we constructed a dataset consisting of 20 phishing emails from Nazario and 20 legitimate emails from Enron. The results presented in Table \ref{tab:llm_performance_on_explanability} highlighted the trade-off between classification accuracy and model's explanation and self-consistency across different LLM model architectures using the baseline models.
Llama 7B and Llama 8B model exhibited high CC-SHAP scores suggesting stronger self-consistency in their prediction-explanation token alignment with $0.9659 \pm 0.030$ for phishing, $0.977 \pm 0.0169$ for ham in Llama 7B and $0.9549 \pm 0.052$ for phishing and $0.974 \pm0.013$ for ham in Llama 8B. However, their phishing accuracy remained low $40$ and $30$ for Llama 7B and Llama 8B. This suggest that while these models may be consistent in their prediction-explanation token alignment, internal faithfulness, they often struggle to differentiate phishing classification effectively. In contrast, Wizard 7B demonstrated a lower CC-SHAP score for phishing $0.123 \pm 0.0906$ and ham emails $0.1925  \pm 0.123$, yet attained a higher phishing accuracy of $80$. This can be implied that Wizard 7B implements the different approach in decision making, and shows lower consistency in their token alignment, prioritizing detection with higher accuracy. Meanwhile, all models performed well in classifying ham emails, with Llama models obtaining $100$ accuracy and Wizard 7B $95$. This finding suggest that there might be a possible bias towards legitimate emails, potentially introducing the false negatives (False Alarm) in phishing classification.

\begin{figure*}
    \centering
    \includegraphics[width=1.0\linewidth]{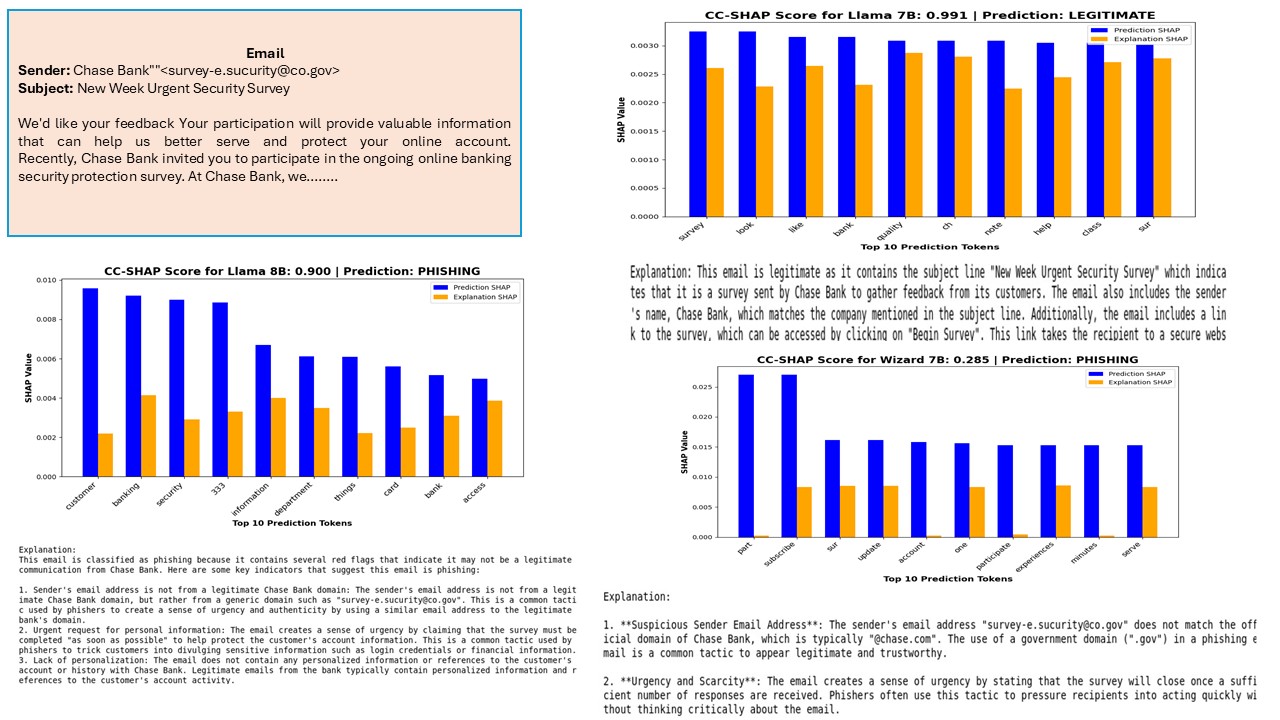}
    \caption{Snapshot of Model Input, Generated Explanation, and CC-SHAP Scores, with SHAP Values for Top Contributing Tokens During Prediction and Their Corresponding Values in the Explanation}
    \label{fig:CC-SHAP Score with the model Prediction and Explanation}
\end{figure*}


\begin{table*}
\centering
\begin{tabular}{|l|c|c|c|c|}
\hline
\textbf{Model} & \textbf{Phishing CC‑SHAP (Mean ± Std Dev)} & \textbf{Ham CC‑SHAP (Mean ± Std Dev)} & \textbf{Phishing Accuracy (\%)} & \textbf{Ham Accuracy (\%)} \\
\hline
LLaMA 7B  & 0.9659 ± 0.0304 & 0.9779 ± 0.0169 & 40.0 & 100.0 \\ \hline
LLaMA 8B  & 0.9549 ± 0.0523 & 0.9742 ± 0.0137 & 30.0 & 100.0 \\ \hline
Wizard 7B & 0.1231 ± 0.0906 & 0.1924 ± 0.1283 & 80.0 & 95.0 \\ \hline
\end{tabular}
\caption{Average CC-SHAP score for emails with their prediction accuracy
\\ Higher the CC-SHAP values, better the tokens align}

\label{tab:llm_performance_on_explanability}

\end{table*}

To gain a better understanding of how different models interpret phishing emails, we applied CC-SHAP explanability and consistency analysis to compare their classification ability and reasoning style across different LLM models using a phishing email as an input as shown in Figure \ref{fig:CC-SHAP Score with the model Prediction and Explanation}. It can be observed that two of these models Llama 8B and Wizard 7B predicted as phishing, while Llama 7B predicted as legitimate for the same email. Along with these prediction, the explanation from each of these models are very different, each prioritizing different features as an major reason for their decisions, demonstrating differences in their underlying decision processes.
The CC-SHAP score from different models varies even for same email ranging from  $0.991$ for Llama 7B and $0.900$ for Llama 8B to $0.285$ for Wizard 7B. The variance in CC-SHAP scores implies how model perceives and weighs different semantic and contextual cues while making classification and explanation reasoning. The different models identified different high-influencing tokens for their predictions and explanations. In case of Llama 7B and Llama 8B words such as ``bank", ``card" , ``protect", "quality", and ``security" were prioritized and which aligns with phishing tactics which creates urgency. Likewise, Wizard 7B focused on ``participate", ``subscribe" and ``update", as an indication of different weighting approach.

Based on the interpreted results, the findings align with our previous observation that high cc-shap score does not always correlate with accurate decision making.

\section{Conclusion}


This research study provides a comprehensive evaluation of LLM models highlighting the challenges in phishing detection, particularly in their explanability, consistency, and classification. We systematically evaluated the efficacy of three fine-tuning techniques: Binary Classification, Contrastive Learning, and Direct Preference Optimization (DPO), across multiple transformer‐based architectures (BERT, LLaMA 7B, LLaMA 8B, and Wizard 7B) on combined Nazario and Enron phishing datasets. We found that binary classification was by far the most reliable approach, achieving the highest accuracy and lowest loss,whereas Wizard 7B lagged in generalization. Similarly, Contrastive learning demonstrates minimal losses for BERT and  for LLaMA 8B,yet delivers no accuracy gain over binary classification. DPO fine-tuning incurs very high losses for LLaMA 7B and for Wizard 7B and underperformed in phishing detection. These results show that rich embedding objectives alone must be paired with explicit classification supervision for reliable performance. In our  CC-SHAP analysis, it is  observed that LLaMA models score high on explanation-prediction token alignment ($>0.95$),  yet low on phishing accuracy (~30\% and ~40\%) , whereas Wizard has lower token alignment  but higher accuracy of ~80\%  highlighting that explanation confidence alone doesn’t predict real‐world performance.

Our findings raise two crucial questions: ``Should LLMs be designed to replicate human-like uncertainty, adopting the complexities of human decision-making? Or is it better they prioritize deterministic behavior to ensure consistency, reliability, and predictability? " \cite{ma2024can}. With these questions as part of future direction, research should be more focused on developing LLMs that balance human-like uncertainty with reliability, ensuring both consistency and adaptability in decision making. Moreover, exploration on human-centric fine-tuning techniques and dynamic integration of LLMs and Cognitive models with improved evaluation metrics can help to understand the human perspective. This approach can facilitate more individualized training modules for phishing detection in the future.

\bibliographystyle{IEEEtran}
\bibliography{reference}
\end{document}